%% file: main.tex
\documentclass[sigconf]{acmart}
\AtBeginDocument{%
  }

\copyrightyear{2022}
\acmYear{2022}
\setcopyright{rightsretained}
\acmConference[UIST '22 Adjunct]{The Adjunct Publication of the 35th
Annual ACM Symposium on User Interface Software and
Technology}{October 29-November 2, 2022}{Bend, OR, USA}
\acmBooktitle{The Adjunct Publication of the 35th Annual ACM
Symposium on User Interface Software and Technology (UIST '22 Adjunct),
October 29-November 2, 2022, Bend, OR, USA}
\acmDOI{10.1145/3526114.3558529}
\acmISBN{978-1-4503-9321-8/22/10}

\input{contents/heading}

\begin{document}

\title{Exploiting and Guiding User Interaction in Interactive Machine Teaching}

\author{Zhongyi Zhou}
\affiliation{
  \institution{Interactive Intelligent Systems Lab., The University of Tokyo}
  \city{Tokyo}
  \country{Japan}
}
\email{zhongyi@iis-lab.org}

\renewcommand{\shortauthors}{Zhou}

\begin{abstract}
% My Ph.D. research studies Interactive Machine Teaching (IMT)
% The modern society gains considerable benefit from many AI applications.
% To develop such applications, researchers typically teach AI new concepts by repetitive labeling.
% In reality, labeling is only one of the vast interactions that humans perform when teaching concepts to human learners.

% Existing work primarily teaches an AI system by repetitive labeling, which is only one of the massive interactions humans perform in teaching.
% Despite the rich information embedded in humans' teaching interactions, 
Humans are talented with the ability to perform diverse interactions in the teaching process.
However, when humans want to teach AI, existing interactive systems only allow humans to perform repetitive labeling, causing an unsatisfactory teaching experience. 
My Ph.D. research studies Interactive Machine Teaching (IMT), an emerging field of HCI research that aims to enhance humans' teaching experience in the AI creation process.
My research builds IMT systems that exploit and guide user interaction and shows that such in-depth integration of human interaction can benefit both AI models and user experience.
\end{abstract}





% comment out when submitting
\settopmatter{printfolios=true}

\maketitle

\input{contents/intro}

\input{contents/saliency}
\input{contents/lookhere}
\input{contents/ongoing}

% \input{contents/conclusion}
\input{contents/discussion}
\bibliographystyle{ACM-Reference-Format}
\bibliography{contents/reference}

\end{document}

%% file: contents/heading.tex
\usepackage{xspace}
\usepackage{bbm}
\usepackage{amsmath}
\usepackage{svg}
\usepackage{xcolor,cancel}
\usepackage{mathtools}
\usepackage{dsfont}
\usepackage{wrapfig}

\usepackage{caption}
\usepackage{subcaption}
\usepackage{eucal}

\usepackage{array}

\makeatletter
\def\Hline{
  \noalign{\ifnum0=`}\fi\hrule \@height 4.\arrayrulewidth \futurelet
  \reserved@a\@xhline}
\makeatother

\usepackage{amsmath}
\usepackage{graphicx}
\usepackage{here}
\PassOptionsToPackage{hyphens}{url}
\usepackage{hyperref} %% for hypertext
\usepackage{url} %これでURLがいい感じに折り返される
\usepackage{textcomp} %% for registared mark　→®
\usepackage{multicol}
\usepackage{subcaption}
\usepackage{comment}
\usepackage{booktabs,dcolumn}
\usepackage{multirow}
\usepackage{siunitx} %% SI単位系の出力
\usepackage{fancyhdr}			% for use Pagestyle
\usepackage{totpages}
\usepackage{arydshln}
\usepackage{enumitem}
\usepackage{svg}
\usepackage{rotating}
\def\etal{\textit{et al.}}

\def\datasetname{\textit{HuTics}\xspace}

\def\vimt{V-IMT\xspace}

\renewenvironment{quote}[1][0.04\linewidth]
  {\list{}{\leftmargin=#1\rightmargin=#1}\item\relax}{\endlist}

\usepackage{color}
\definecolor{orange}{RGB}{255,127,0}
\definecolor{darkgreen}{RGB}{0, 146, 0}
\definecolor{violet}{RGB}{148,0,211}

\newcolumntype{L}[1]{>{\raggedright\let\newline\\\arraybackslash\hspace{0pt}}m{#1}}
\newcolumntype{C}[1]{>{\centering\let\newline\\\arraybackslash\hspace{0pt}}m{#1}}
\newcolumntype{R}[1]{>{\raggedleft\let\newline\\\arraybackslash\hspace{0pt}}m{#1}}

\newif\ifCOMMENTS
\COMMENTStrue
\ifCOMMENTS

\else

\fi

\makeatletter
\def\Hline{
  \noalign{\ifnum0=`}\fi\hrule \@height 4.\arrayrulewidth \futurelet
   \reserved@a\@xhline}
\makeatother

%% file: contents/intro.tex
\section{Introduction}
% \begin{itemize}
%     \item explanatory feedback to enhance humans' interpretation of ML models
%     \item in-depth human understanding that achieve in-situ annotation
%     \item in-situ feedback to guide diverse teaching
% \end{itemize}
% \begin{itemize}
%     \item IMT introduction
%     \item existing work primarily allows human teachers to teach AI new concepts via repetitive labeling.
%     \item When human teachers teach human learners, the teachers have the ability to incorporate various interactions into the teaching event during the human-human communication.
%     \item Such interactions can deli-ever teaching concepts more efficiently than teaching via simple labeling. For example, \xxx
% \end{itemize}
Artificial Intelligence (AI) is changing the world by assisting various applications that benefit humans' life~\cite{zhou2021syncup, zhong2022bringing, hayashi2021vision}.
To create new AI applications, researchers usually teach AI new concepts~\cite{zhu2015machine} (e.g., what a cat looks like) by training Machine Learning (ML) models~\cite{simonyan2014vgg, vaswani2017attention} on large-scale datasets labeled by human annotators~\cite{imagenet}.
Despite the importance of data labeling, I argue that it is only one of the massive interactions humans are talented at performing during the teaching process.
Humans' teaching interaction provide rich information about the concept that humans want to teach, and thus such interaction should be exploited and encouraged when humans interact with AI~\cite{sultanum2020Teaching}.
However, due to the lack of interactive technologies that support diverse teaching behaviors, existing systems mainly constrained human teachers in performing repetitive labeling, causing an unsatisfactory teaching experience.

Interactive Machine Teaching (IMT)~\cite{ramos2020imt, simard2017machine} is an emerging field of research in HCI that aims to enhance humans' teaching experience during the creation of Machine Learning (ML) models.
Different from developing models through data collection and programming that require professional expertise, typical Vision-based IMT (\vimt) systems allow users to teach a model by demonstrations, which are intuitive behaviors people perform in teaching.
For example, Teachable Machine~\cite{Carney2020teachable} allows users to teach vision-based ML classifiers by demonstrating the objects of different classes in front of a camera.
After the teaching process, the system automates all machine learning processes, and the user can further assess the created model, as well as decide whether they need to perform another iteration of teaching-and-assessing processes~\cite{dudley2018review}.
Based on such standard system designs, recent studies show that IMT systems should further engage non-experts by providing guidance~\cite{smith-renner2020no, Hong2020Crowdsourcing} as well as exploiting more human interactions beyond labeling~\cite{ramos2020imt, sultanum2020Teaching} during the teaching process. 
For example, Fiebrink~\etal~\cite{Fiebrink2011Human} found that users wanted more information in the assessment interface so that they could understand \textit{``where and how the model was likely to make mistakes''}.
Hong~\etal~\cite{Hong2020Crowdsourcing} further highlighted the importance of guidance in supporting users to decide \textit{``what to show in the teaching set''}.

My Ph.D. research focuses on exploiting and guiding users' teaching interaction in \vimt systems.
My first project provided teaching guidance by enhancing users' interpretation of the trained models.
I created an assessment interface with saliency map visualization that explains what portions of the images the model weighs heavily in the prediction.
I further found that the model created by standard \vimt systems may easily misinterpret a concept by highlighting unrelated features.
My second project summarized the cause of the issue as a lack of fine-grained annotations of objects of interest that users want to teach.
To address this issue, I created a \vimt system, called LookHere, that integrates object annotations into the teaching process by exploiting users' deictic gestures towards objects of interest.
The user study shows that the in-situ object annotation achieved by exploiting humans' gestural interaction can significantly accelerate the teaching process without a noticeable model accuracy drop.
In addition to exploiting user interaction, in my third (on-going) project, I propose a \vimt that guides users to perform informative teaching.
The teaching interface will visualize how different a given view can be from the existing teaching set in real time, and thus encourage users to cover a wide range of views in the teaching set.

%% file: contents/saliency.tex
\section{Model Assessment via Saliency Maps}
\begin{figure}
% \centering
    \begin{subfigure}[b]{\linewidth}
        \centering
        \includegraphics[width=\linewidth]{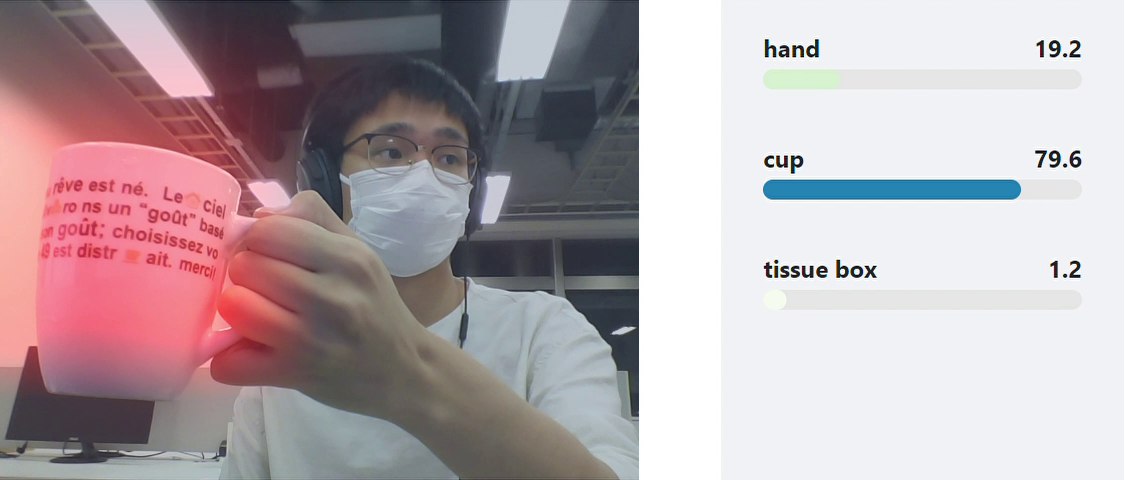}
        \caption{The correct prediction with accurate highlights.}
        \label{fig: model_test_correct}
    \end{subfigure}
    \begin{subfigure}[b]{\linewidth}
        \centering
        \includegraphics[width=\linewidth]{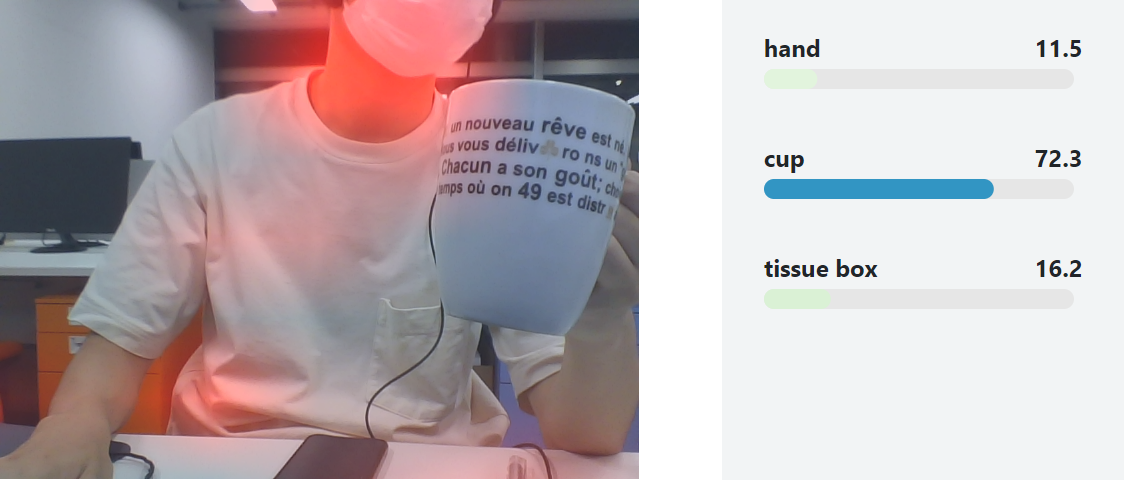}
        \caption{The correct prediction with wrong highlights.}
        \label{fig: model_test_wrong}
    \end{subfigure}
    \caption{Example views of the model testing interface~\cite{zhou2021enhancing}.
    Bar visualization of confidence scores is shown on the right side.
    A camera view with heat-map overlays that explains the model prediction is shown on the left side.
    (a) shows an example with a correct explanation while the explanation in (b) is wrong.  
    }
    \label{fig: model_test}
    \Description[(a) highlights ``cup'', and (b) highlights the human face]{Both figures show prediction of ``cup'' with high confidence. (a) highlights a ``cup'' held by the user in the camera view, and (b) highlights the human face instead of ``cup''.}
\end{figure}
The model assessment interface is a critical component of the IMT system.
Using the interface, the user can be informed of the model performance, which decides whether and how they will teach the model to correct errors.
Existing designs for model assessment mainly present confidence scores of each class as assistant data when the user interacts with the camera.
Despite its usefulness, I argue that there is a lack of support for the user to interpret why the model makes each prediction.
Intuitively, when a human teacher wants to assess whether a human student has fully understood a concept, the teacher may not only care about the student's final answer but also whether the student can answer with appropriate reasons. 
In this work, I enhanced the model assessment interfaces in \vimt through real-time saliency map visualizations.

Figure~\ref{fig: model_test} illustrates an example view of my model assessment interface.
In addition to the confidence scores shown on the right side, the interface also presents saliency maps in real time on the left side.
The saliency maps help the user interpret why the model predicts a target category (the default is the one with the highest confidence score, i.e., ``cup'' in Figure~\ref{fig: model_test_correct}$\&$\ref{fig: model_test_wrong}).
By examining the highlighted regions, the user can validate whether the ML model has correctly understood each concept by visualizing the correct regions.
For example, in Figure~\ref{fig: model_test_correct}, the model provides a correct explanation of what ``cup'' is, whereas the model in Figure~\ref{fig: model_test_wrong} mistakenly highlights the masked human face as the ``cup'', showing the learning failure.
Interestingly, both models show similarly high confidence ($\sim70\%$) for the prediction, implying that the user may trust the model if there is no assistance from my saliency maps.

Such a case when the model is highly confident of a correct prediction but uses a totally wrong reference is not rare.
This reveals a potential risk in prior \vimt systems: the model created from the simplified teaching experience may not precisely learn a target concept.
The main reason is that when the user demonstrates the object in front of a camera, the camera view may inevitably include other visual data than the target object.
Therefore, what the user defines as, for instance, ``cup'', is the full scenes of the images instead of portions of the images that represent ``cup''.
This reveals a teaching ambiguity issue in existing \vimt systems, and future work should investigate how to support the user better clarify the object to teach.
% As a result, even though the
% For example, if the user teaches ``cup'' by
% object that s/he wants to teach,
% non-ML-experts want to teach ML
% show the  by highlighting the image regions that the created model weighs heavily. 
% Such visualization serves as an addition verification approach in the model testing stage, allowing the user to confirm whether the model has mistakenly interpret other unrelated contents as the target object.
% For example, in Figure~
% On the right side of the view, there are three bar plots which visualize the real-time confidence scores 
% \begin{itemize}
%     \item It is not sufficient to perform model testing only via confidence scores
%     \item I created a model testing tool that visualizes the region of interest that the trained model uses for each prediction
%     \item My exploration reveals that the saliency maps can expand the evaluation dimension in the model testing stage. I also find that the model may highlight unrelated features of the taught object.
% \end{itemize}

% \xxx some major findings

% \subsection{Major Findings}
% 

%% file: contents/lookhere.tex
\section{Gesture-aware Teaching with In-situ Annotations}
% \begin{itemize}
%     \item Motivation: neither neglecting annotations nor post-hoc annotations works well in IMT scenarios.
%     \item LookHere Teaching Interface
%     \item gesture-aware object-agnostic segmentation
%     \item hutics
% \end{itemize}

% The study in the previous section shows that the model created through simplified teaching processes may be unreliable because it may misinterpret a target object using unrelated features.
One na\"ive approach to address the teaching ambiguity issue is to let the user annotate the object of interest in each teaching sample, emphasizing which object the model should learn.
Although existing research on interactive annotation provided simplified interaction for the annotation process (e.g., by clicking~\cite{mortensen1998interactive, fbrs2020sofiiuk}, sketching~\cite{rother2004grabcut, Zhang2020weakly}, or mouse dragging~\cite{chang2021spatial}, such post-hoc annotations inevitably bring extra workload that may diminish human teachers' user experiences.

Can \vimt systems integrate object annotations into teaching so that the user can express in-situ annotations through simple interaction?
In this project, I exploited humans' deictic gestures for achieving such integration and built a gesture-aware \vimt system, called LookHere.
The main intuition behind the system design is that humans can naturally interact with the object using deictic gestures in the teaching process.
For example, during the teaching process, the user may hold the object or point at the object.
Such gestures explicitly provide important cues on where the object of interest is.
My system, LookHere, can intelligently capture such cues embedded with humans deictic gestures and achieves in-situ annotations in the teaching process.
% \xxx

\subsection{LookHere}
\begin{figure}
    \centering
    \includegraphics[width=\linewidth]{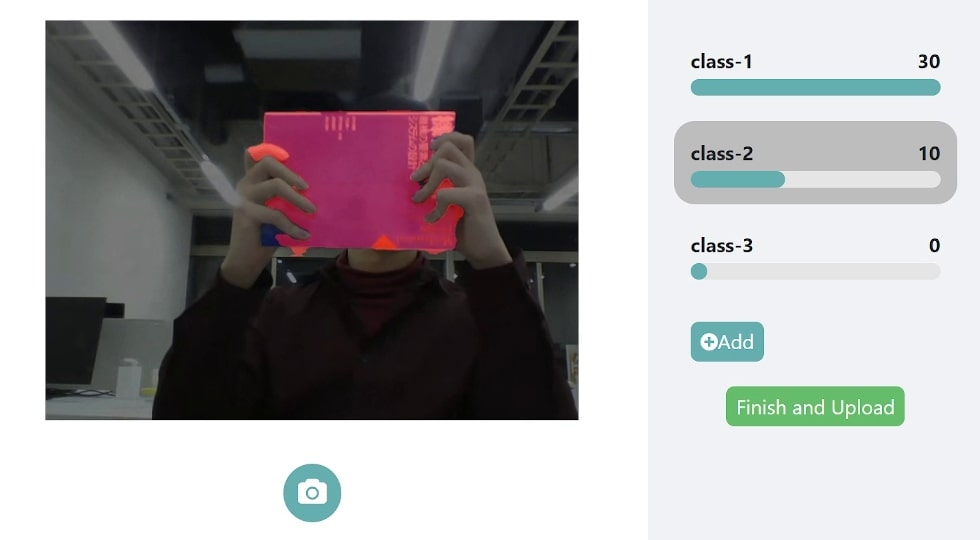}
    \caption{LookHere~\cite{zhou2022gesture}. The camera view with object highlights is presented on the left side of the visualization of the count of teaching samples in each category.}
    \label{fig: model_teach}
    \Description[The book held by the user is highlighted in red.]{The user is teaching LookHere by holding a book. The book is highlighted in red.}
\end{figure}

The key characteristic of LookHere is that it integrates the annotation process into human teaching by leveraging the user's deictic gestures.
Figure~\ref{fig: model_teach} shows an example view of the teaching interface in LookHere.
Other than standard functions in other \vimt systems~\cite{Carney2020teachable, francoise2021marcelle} (e.g., visualizing counts of teaching samples in each category), LookHere includes a function called \textit{object highlights} to inform the user what regions of the camera view the system considers as the object to learn.
The object highlights work in real time at approximately 28.3 fps using one GTX 2080Ti GPU.
% LookHere achieves this using our gesture-aware object segmentation algorithm, which will be explained in detail in the next subsection.
After the user clicks on the camera icon on the bottom-left corner of Figure~\ref{fig: model_teach}, LookHere saves an image-label pair, as well as a segmentation mask which is visualized as object highlights on the teaching interface. 

% The additional segmentation mask saved in real time with the image-label pair resembles a annotation process but it does not require th
% an additional segmentation mask of the object in addition to the , a common data format in prior \vimt systems.
% The additional segmentation mask can better supervise the training process of the model to be created.
% As is shown in Figure~\ref{xxx}, in confusing situations, the user can teach (i.e., highlighted region) by performing different gestures towards the target object.

\subsection{Gesture-aware Object Highlights}
\begin{figure}
\centering
\includegraphics[width=\linewidth]{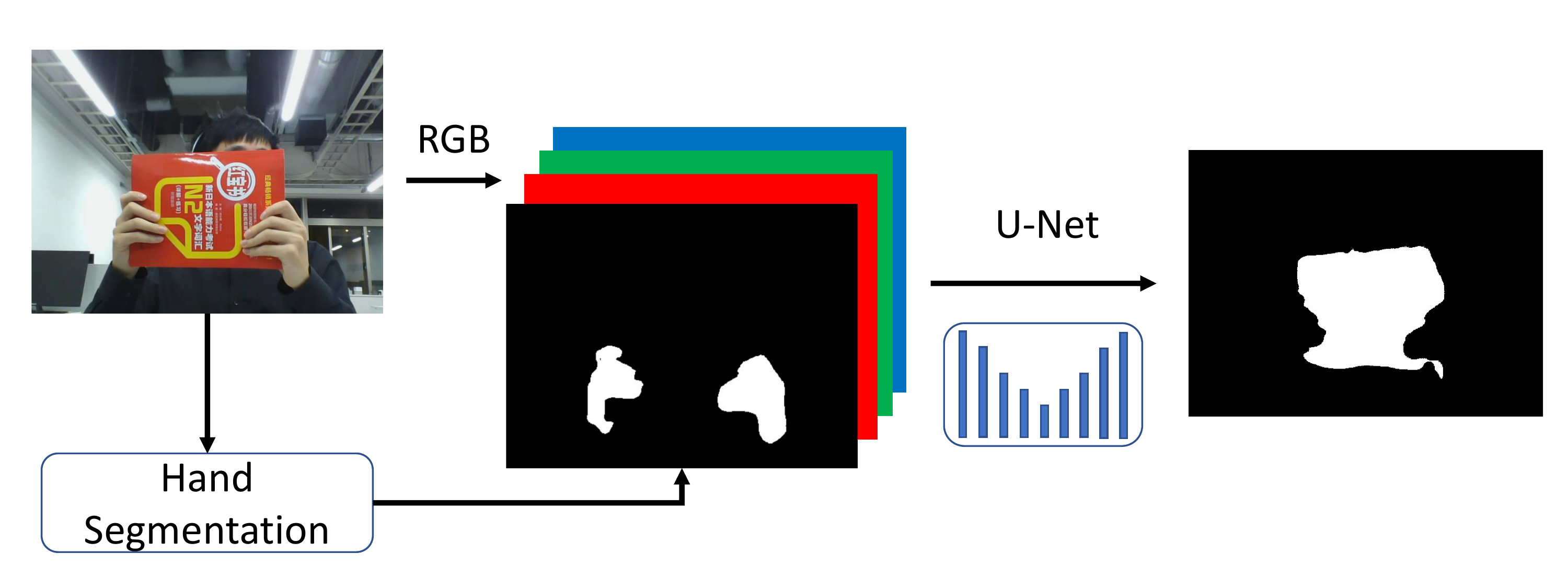}
\captionof{figure}{The workflow of the gesture-aware object-agnostic object segmentation algorithm~\cite{zhou2022gesture}. 
The object highlights are created by using a U-Net and inputting the original RGB image attached to its hand segmentation result.
}
\label{fig:implementation workflow}
\Description[This is a workflow figure. Given an image in which a person is holding a book, the workflow first predicts a hand segmentation mask and then outputs the segmentation of the book.]{This figure illustrates the process to generate a visual overlay for the object guided by users’ deictic gestures. The image taken from a camera shows that a person is holding a book with both hands in this example. Through the hand segmentation method, the system creates a hand segmentation mask in addition to the original RGB images. Through U-Net, the system predicts the region of the object guided by users’ deictic gestures. This prediction is the output of this process and leads to the segmentation mask of the object.}

\end{figure}

LookHere achieves the in-situ annotation using our gesture-aware object segmentation algorithm.
To support real uses in \vimt, the segmentation algorithm ought to be object-agnostic (i.e., without the constraints of what object can be segmented) because \vimt systems allow users to teach a wide range of daily objects.
LookHere addresses this challenge through a gesture-aware algorithm as well as our customized dataset.
Figure~\ref{fig:implementation workflow} summarizes the workflow of our algorithm.
LookHere first performs a hand segmentation with a given RGB image.
It can then feeds both the RGB image and the hand segmentation mask into U-Net, which predicts a segmentation mask of the object specified by deictic gestures.
The main intuition of algorithm design is that the hand segmentation should provide informative cues on where the object specified by the users' gestures is, making it possible to perform object-agnostic segmentation.

\subsection{HuTics}
% There is no existing datasets that are collected for my target purpose. 
The most related dataset to the use case of LookHere is TEgO~\cite{Lee2019Hands}, and thus I tried training the algorithm on TEgO with augmented labels. 
The results show that the trained model is not robust enough to accurately various daily objects.
Therefore, in this work, I collected a customized dataset, called HuTics, which included 2040 images collected from 170 people using human deictic gestures to interact with diverse daily-life objects.
To cover a wide range of deictic gestures that humans may perform, I refer to Sauppe et al.’s taxonomy~\cite{sauppe2014robdeic} and divide deictic gestures into four categories (i.e., exhibiting, pointing, presenting and touching).
Each participant was required to use these four kinds of deictic gestures to interact with daily-life objects and took 12 photos in total (i.e., three photos for each kind of deictic gesture).
I highly encouraged them to interact with a wide range of daily-life objects in their 12 photos.
I then recruited five annotators to label the segmentation of the object.

We further used the data from 80\% of the participants in HuTics (i.e., 1632
images from 136 people) for training and 20\% for testing. 
% Using U-Net~\cite{ronneberger2015unet} with EfficientNet-b0~\cite{tan2019efficientnet} backbone, the $mIoU$ accuracy is $0.718$, and it can be run at 28.3 fps using one GTX-2080Ti GPU.
% In comparison, the $mIoU$ accuracy of the same network trained on TEgO with augmentation is $0.368$, showing a big accuracy drop.
We found that the accuracy of our gesture-aware algorithm (Figure~\ref{fig:implementation workflow}) trained on HuTics is $0.718$.
The accuracy of the same algorithm trained on TEgO with augmentation is $0.368$, showing a large accuracy drop.
Please see our full paper for details of the algorithm choice~\cite{zhou2022gesture}.
This demonstrates that HuTics plays a critical role in powering my object-agnostic segmentation algorithm compared to directly using other related datasets with augmentation.

\begin{figure*}[t]
    \centering
    \begin{subfigure}[b]{0.495\linewidth} 
        \centering
        \includegraphics[width=\textwidth]{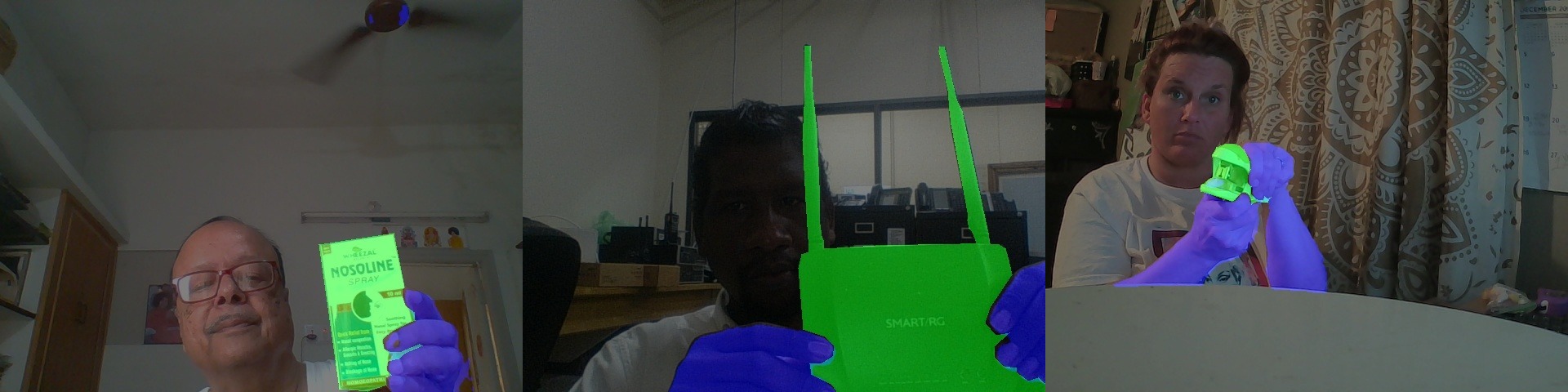}
    \end{subfigure}
    % \\
    \begin{subfigure}[b]{0.495\linewidth}
        \centering
        \includegraphics[width=\textwidth]{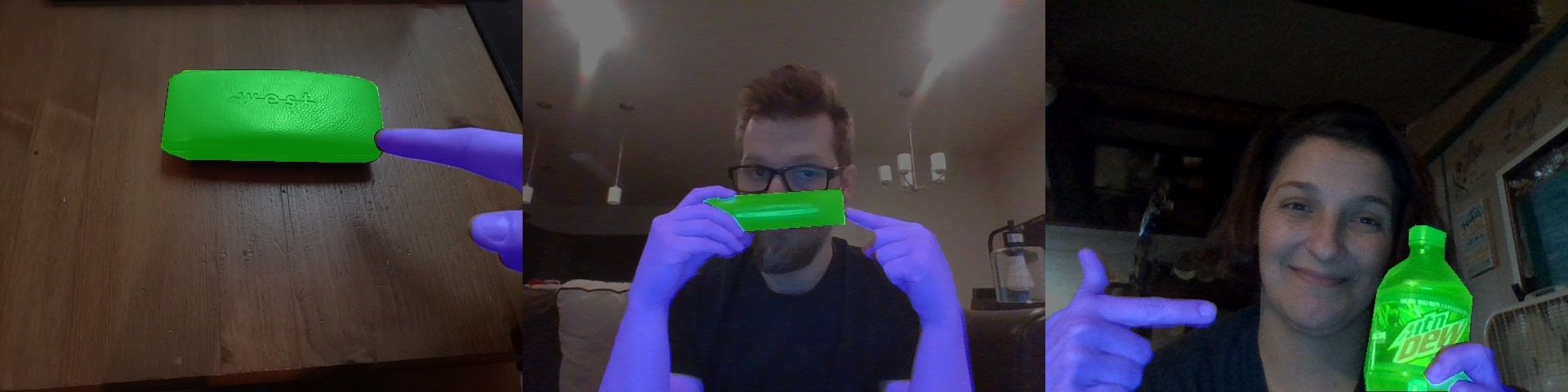}
    \end{subfigure}
    \begin{subfigure}[b]{0.495\linewidth} 
        \centering
        \includegraphics[width=\textwidth]{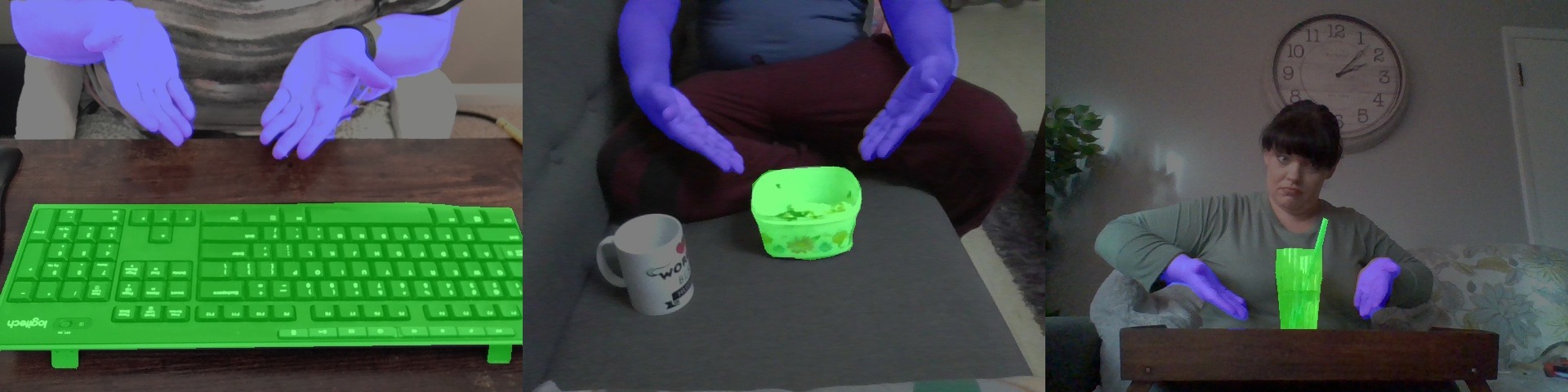}
    \end{subfigure}
    % \\
    \begin{subfigure}[b]{0.495\linewidth}
        \centering
        \includegraphics[width=\textwidth]{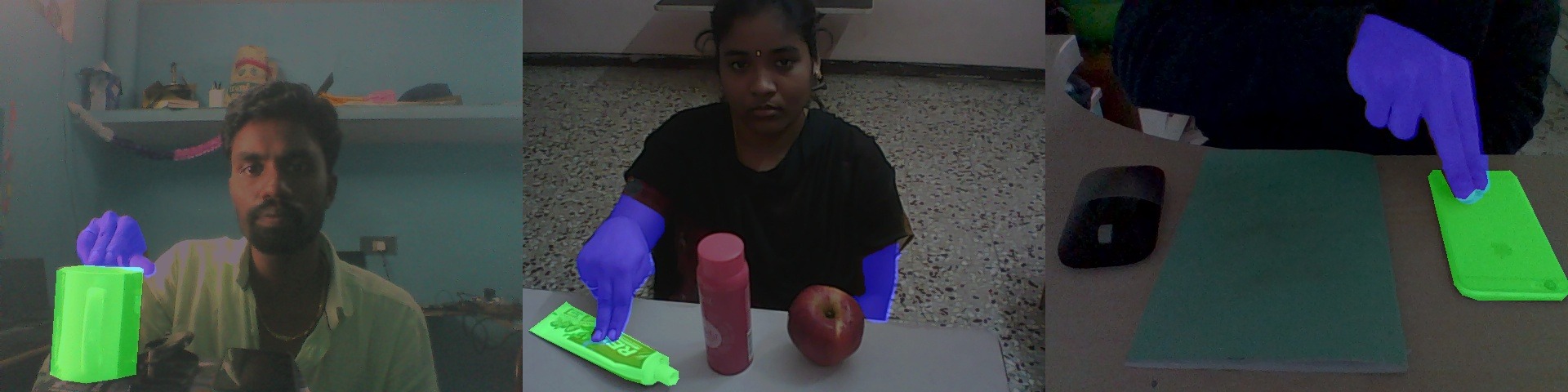}
    \end{subfigure}
    \caption{Example images with visualization in \datasetname dataset. 
    \datasetname covers four kinds of deictic gestures to objects: exhibiting (top-left), pointing (top-right), presenting (bottom-left) and touching (bottom-right).
    The hands and objects of interest are highlighted in blue and green, respectively.
    Note that there is no human annotation of hand segmentation, and the blue regions are from the prediction of Li et al.'s method~\cite{Li2020self}.
    % Top-left: exhibiting. Top-right: pointing. Bottom-left: presenting. Bottom-right: touching. 
    % \koji{}{it's better to avoid using pictures that include faces (even if we have approvals from them).}
    }
    \label{fig: hutics} 
    \Description[This figure includes visualizations of how different users specify an object using gestures.]{This figure includes a $2\times 2$ grid of subfigures. In each subfigure, there are three images. In each image, a person specifies an object using gestures.}

\end{figure*}

\subsection{Evaluations}
To evaluate LookHere, we conducted a user study with 12 participants.
None of them has experience in studying or working in the fields related to AI or ML.
The purpose of this study is to understand the benefits of LookHere by comparing it with three baseline interface designs:
\setlength{\leftmargini}{10pt}
\begin{itemize}
\item \textit{Na\"{i}veIMT}:
This represents the most common design in current \vimt systems~\cite{Carney2020teachable, francoise2021marcelle}, in which participants only perform object demonstration without annotation.
\item \textit{Contour}:
In addition to the teaching process with the na\"{i}ve IMT system, this condition would involve a manual annotation procedure, using a contour-based tool~\cite{bounias2021interactive}, in a post-hoc manner.
\item \textit{Click}: 
This condition replaces the contour-based tool in the ``Contour'' condition with a click-based annotation method~\cite{sofiiuk2021reviving} to represent a simplified annotation process.
\end{itemize}
Participants were required to teach a vision-based classification model under each given interface condition, and we measured three kinds of metrics in the study: 1) time consumption; 2) model accuracy; 3) NASA-TLX~\cite{hart1988development}. 
Note that the model accuracy includes both classification accuracy as well as segmentation accuracy, aiming to test whether the model can not only predict correctly but also explain the prediction accurately.

\begin{table}[t]
\caption{The mean values and standard deviations of the task completion time and accuracies (classification and segmentation) across the four interface conditions.
}
\label{tab: res_time_acc}
\begin{tabular}{cc|c|c|c|c}
\hline
\multicolumn{2}{c|}{}                             & LookHere                                                & Na\"{i}veIMT                                                & Click                                                   & Contour                                                 \\ \hline
\multicolumn{2}{c|}{Time {[}s{]}}                 & \begin{tabular}[c]{@{}c@{}}104\\ (44)\end{tabular}      & \begin{tabular}[c]{@{}c@{}}67\\ (10)\end{tabular}       & \begin{tabular}[c]{@{}c@{}}1,197\\ (228)\end{tabular}   & \begin{tabular}[c]{@{}c@{}}1,483\\ (407)\end{tabular}   \\ \hline
\multicolumn{1}{c|}{\multirow{2}{*}{Acc.}} & Cls. & \begin{tabular}[c]{@{}c@{}}0.824\\ (0.158)\end{tabular} & \begin{tabular}[c]{@{}c@{}}0.847\\ (0.190)\end{tabular} & \begin{tabular}[c]{@{}c@{}}0.880\\ (0.141)\end{tabular} & \begin{tabular}[c]{@{}c@{}}0.833\\ (0.159)\end{tabular} \\ \cline{2-6} 
\multicolumn{1}{c|}{}                      & Seg. & \begin{tabular}[c]{@{}c@{}}0.605\\ (0.153)\end{tabular} & \begin{tabular}[c]{@{}c@{}}0.139\\ (0.095)\end{tabular} & \begin{tabular}[c]{@{}c@{}}0.716\\ (0.167)\end{tabular} & \begin{tabular}[c]{@{}c@{}}0.732\\ (0.151)\end{tabular} \\ \hline
\end{tabular}
\end{table}

Table~\ref{tab: res_time_acc} summarizes the results of time consumption and two types of accuracies.
The results reflect that LookHere can enable a fast model creation experience without significantly sacrificing the model accuracy.
NASA-TLX results (see detailed data in the full paper~\cite{zhou2022gesture}) further show LookHere causes a lower cognitive burden compared to two conditions that require post-hoc annotations.
Therefore, our user study demonstrates that LookHere can achieve a good balance between accuracy and workload for \vimt due to our system designs that integrate annotations into teaching.

%% file: contents/ongoing.tex
\section{Teaching Guidance by Visualizing Data Diversity in Real Time}
% LookHere provided real-time feedback on what regions in the camera view the system considers as the objects of interest to learn.

LookHere solves the issue of clarifying \textit{what} (regions in the camera view) the user wants to teach by exploiting gestural interaction.
In addition to exploiting human interaction, I argue that IMT research should also motivate human interaction by guiding users \textit{how} to perform informative teaching.
In practice, it is highly challenging to guide users' teaching behaviors in real time because it usually takes time for a system to judge whether the data are valuable for training or not.
To compute the data value~\cite{yoon2020data}, a system needs to train a back-end model on the data, which is a very time-consuming procedure, particularly for deep learning models.
Fails and Olsen~\cite{fails2003interactive} also highlighted this issue when they first introduced interactive machine learning research.
They further argued that researchers should use lightweight models like decision trees (DTs) as the back-end models instead of Neural Networks (NNs) so that users can receive rapid feedback on their teaching behaviors.

% can contribute to a high-accuracy model.
% there is no straight-forward method to compute the value of teaching data.

% The main reason is that the system needs to quantify the value of the teaching data by evaluating the model trained on the data, and the training process usually is very time-consuming.
% train the model using these data so that the system can quantify the value of the teaching data by computing the accuracy of the trained model.
% To achieve rapid training process for quick feedback to users,

% This work considers
% I further derive the following research question to be answered in this project:
% \begin{itemize}
%     \item \textit{How can a \vimt system provide feedback on the value of the teaching data are without training the back-end model on the dataset?}
% \end{itemize}
% By exploring this research question, I aim to create a \vimt system that guides users to perform informative demonstration in front of the camera during the teaching process.

% More importantly, its computation is 
% Therefore, instead of training a model on the data, I propose a data valuation method for real-time \vimt teaching guidance by calculating data diversity in a given teaching data.
% Such computation causes much lower burden compared to the optimization process in machine learning, and therefore is feasible to be used for the purpose of real-time feedback.
% It also brings the 
In this on-going project, I plan to create a system that provides real-time teaching guidance when the user is teaching a deep learning model.
To achieve this goal, this study challenges a widely acknowledged design principle that the system provides feedback \textit{after} the model finishes training on the full training set~\cite{zhou2021enhancing, Carney2020teachable}.
Instead, I argue that the quality of teaching data can be computed \textit{before} the time-consuming training process, which overcomes the main bottleneck of rapid feedback designs in IMT research.
This work considers data diversity as a key factor that can boost the informativeness of teaching, which is a commonly acknowledged heuristic that can benefit ML models~\cite{kang2021styling}.
Compared to the data value, data diversity is much easier to compute.
% More importantly, the computation of data diversity is independent of the choice of back-end models, making it possible for the real-time diversity feedback to be accessible to the IMT system that trains deep learning models at the back end.
More importantly, the computation of data diversity is independent of the choice of back-end models, making it possible for the real-time feedback in IMT system that trains deep models at the back end.

% \zhongyi{}{revise this sentence}

\begin{figure}
% \centering
    \begin{subfigure}[b]{0.49\linewidth}
        \centering
        \includegraphics[width=\linewidth]{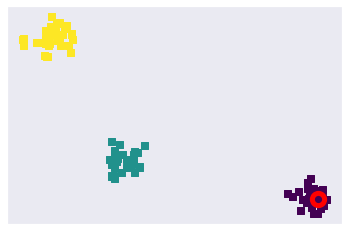}
        \caption{Low data diversity.}
        \label{fig: low}
    \end{subfigure}
    \begin{subfigure}[b]{0.49\linewidth}
        \centering
        \includegraphics[width=\linewidth]{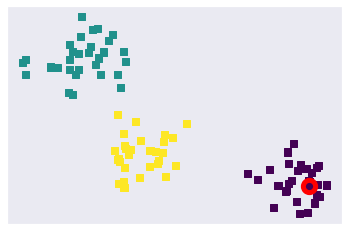}
        \caption{High data diversity.}
        \label{fig: high}
    \end{subfigure}
    \caption{
    Example views of visualization in the teaching interface.
    (b) describes a distribution of teaching data that has higher data diversity than (a).
    }
    \label{fig: diversity}
    \Description{In each subfigure, there are three clusters. The clusters in (b) are more sparse than those in (a).}
\end{figure}

Figure~\ref{fig: diversity} illustrates two examples of how I plan to visualize the data diversity in the teaching interface when the user teaches a three-way classifier.
Each rectangle in the visualization represents an image recorded in the system, and the color represents its classification label. 
The circle highlighted by the red edge is the image that the user is currently teaching, which moves in real time with the change of the camera view.
By continuously interacting with the circle, the user is encouraged to build each cluster to be as large and sparse as possible (i.e., Figure~\ref{fig: high} is better than Figure~\ref{fig: low}).
For example, both circles in Figure~\ref{fig: high}$\&$\ref{fig: low} are not ideal teaching data to be added to the teaching set since both fail to expand the coverage of the purple categories. 
% \zhongyi{}{Revise}
I envision that such real-time visualization can motivate the users to present diverse views of the target objects, benefiting the machine learning process.  
% The circle highlighted by the red edge moves
% indicates the position of the teaching datum if the user records the current moment as a 
% I envision that such visualization
% The user is encouraged to interact with the visualization so that the system can collect a wide range of data within each category.
% feedback on the quality of the teaching data by calculating its degree of diversity, which can be computed in real time regardless of the size of the model to be trained.
% Specifically, I use
% In addition, I plan to provide visualization of the data distribution in the semantic layer, which encourages to perform teaching so that the teaching data within each category can be spanded

% instead of providing feedback based on the trained model of the full dataset, this work creates an IMT system that provides immediate feedback that estimate the goodness of the teaching samples by the diversity of teaching data.

% To implement t
% I further investigate methods to provide feedback BEFORE the training process, and even BEFORE people finishes teaching the whole dataset.

% To achieve the useful fee

% go through a training process on these data, which is highly time-consuming and almost impossible to finish in a short period of time.

%% file: contents/discussion.tex
\section{Discussion and Future Work}
User interaction beyond labeling during the teaching process require more in-depth studies to benefit both user experience and AI models.
My research studies user interaction in IMT systems by exploiting deictic gestures and guiding users' object demonstration processes.
In reality, humans use many interactions in the teaching process, and therefore more user interactions than those covered by this paper should be exploited and guided to enhance human-AI collaboration.
For example, future work can study how to exploit gaze and verbal interaction that also contain rich information on the concepts users want to teach.
On the other hand, users still encounter many challenges in which they have no idea how to perform effective teaching~\cite{Hong2020Crowdsourcing, Yin2019Understand}.
Future work should study how to guide user interaction in these challenging scenarios (e.g., how systems can support users to correct a model that misinterprets a concept other than simply labeling more data).

It is important to note that this paper assumes that there is only one human teacher in the human-AI interaction.
However, in practice, multiple users may teach models collaboratively.
% , which is another future direction of IMT research.
With more users engaged in the teaching process, new types of teaching interactions may emerge, bringing new research questions that requires investigation.
Future research should observe these interactions and further study how the interactions can be exploited and guided to enhance collaboration.